\documentclass[apj]{emulateapj}
\usepackage[]{natbib}
\bibpunct{(}{)}{;}{a}{}{,}

\slugcomment{Publications of the Astronomical Society of the Pacific, 121:1297-1306, 2009 December}

\shorttitle{Bulgeless Galaxies}
\shortauthors{Stefan J. Kautsch}

\begin{document}

\title{The Edge-On Perspective of Bulgeless, Simple Disk Galaxies}

\author{Stefan J. Kautsch\altaffilmark{1}}

\altaffiltext{1}{Department of Physics, Computer Science and Engineering, Christopher Newport University, Newport News, VA 23606}
\email{stefan.kautsch@cnu.edu}

\begin{abstract}
This review focuses on flat and superthin galaxies. These are edge-on bulgeless galaxies, which are composed of a simple, stellar disk. 
The properties of these simple disks are at the end of a continuum that extends smoothly 
from bulge-dominated disk galaxies to the pure disks. On average, simple disks are low-mass galaxies with low surface brightnesses, blue colors, and slow rotational velocities. 
Widely-accepted cosmological models of galaxy formation and evolution were challenged by a relatively large observed fraction of pure disk galaxies, 
and only very recent models can explain the existence of simple disk galaxies. This makes simple disks an optimal 
galaxy type for the study of galaxy formation in a hierarchical Universe. They enable us to analyze the environmental and internal influence on 
galaxy evolution, to study the stability of the disks, 
and to explain the nature and distribution of dark matter in
galaxies. This review summarizes the current status of edge-on simple disk galaxies in the Universe. 

\end{abstract}


\section{Introduction and History}
After the Great Debate in 1920 \citep{trimble95}, it became evident that many of the known nebulae were extragalactic systems. \citet[][]{hubble26} introduced a classification
scheme for extragalactic nebulae that is still the most powerful tool today to categorize galaxy morphology \citep[see also][]{vdbergh07}. In this scheme, galaxies of different
morphologies can be reduced into two basic geometric manifestations: stellar spheroidal ellipsoid or stellar disk. 
All other morphologies represent a combination of spheroidal
components centered in disks, and span a continuum from the spheroid-dominated early-type galaxies (E, S0, Sa) to the disky late-type galaxies (Sb, Sc, Sd, Sm, Im, Irr).
Peculiar and distorted morphologies are considered to be the result of interaction processes \citep[][]{pfleiderer63,toomre77}.

\begin{figure}
\epsscale{1}
\plotone{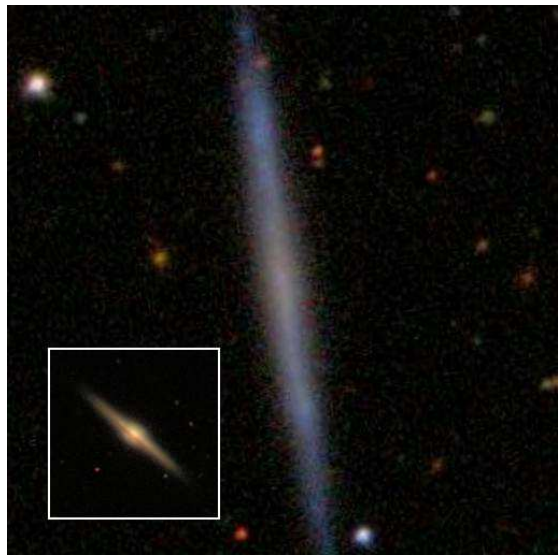}
\caption{The late-type, bulgeless edge-on galaxy J214439.43-064122.5, an Sd(f) class galaxy, is an example of 
a simple disk.~For contrast, the inlay shows a typical disk galaxy (J111146.36+364442.3) with a bulge of type Sa(f). 
Note the typical dust lane in the Sa(f) galaxy which is absent in the simple disk galaxy. Both images are taken from the 
SDSS and shown in \citet{kautsch09a}. The angular size of the images is 100 arcsec$^2$.}
\label{fig1}
\end{figure}

In the 1960s, a special type of a thin and elongated galaxy was found in various galaxy surveys.~\citet{ogorodnikov57,ogorodnikov58} 
and \citet{vorontsov67,vorontsov74} were among the first scientists 
who studied these needle-shaped galaxies. \citet{fujimoto68} suggested that 
these systems are very elongated, prolate ellipsoids. However, the needles would be gravitationally and kinematically 
unstable systems. It later became evident that these objects are bulgeless ``simple disk'' galaxies seen edge-on (\citet{heidmann72}, see also \citet{caimmi07}). Figure~\ref{fig1} 
shows an example of an edge-on simple disk galaxy in contrast to a disk galaxy with bulge. 

Because of their appearance, these galaxies are frequently called flat galaxies \citep[e.g.,][]{karachentsev89,karachentsev93}.~Flat galaxies 
are edge-on disks that are defined to have axial ratios of the semi-major to semi-minor axis of $\geq$ 7 on 
blue photographic plates \citep{karachentsev93,karachentsev99a}; for instance, \object{M33} would be a flat galaxy when seen edge-on. 
Almost all flat galaxies are bulgeless disks. Objects with even larger axial ratios ($\frac{a}{b} \geq 10$) are called superthin 
galaxies and represent a subset of bulgeless flat galaxies with very small disk scale heights \citep{goad79,goad81}. Flat bulgeless 
and superthin galaxies are part of the class of simple disk galaxies, which ranges from thin, late-type galaxies of morphological 
Hubble class $\sim$Scd and later without a bulge component\footnote{Not all late-type spirals are bulgeless \citep{boeker03,graham08}.} to 
the thicker, puffy disks of bulgeless irregular disks. The Large Magellanic Cloud, \object{LMC}, represents a prototype for a non edge-on 
irregular and puffed bulgeless disk \citep{wyse97}. This review focuses on the properties and challenges related to bulgeless flat and 
superthin galaxies as an integral part of the class of simple disk galaxies.

\section{Formation}

The formation of disk galaxies in general is believed to be the result of the collapse of a gaseous protogalaxy within a dark halo \citep{eggen62,white78,fall80}. 
Chemodynamical and analytical models of disk evolution within a slowly growing dark matter (DM) halo can reproduce many properties of disk 
galaxies like the Milky Way \citep{samland03,hernandez06,dutton09}. In these models, 
a Gaussian distribution of initial conditions leads to either a massive disk galaxy after 
an efficient collapse of a
low angular momentum protogalaxy or to a low surface-brightness (LSB) exponential disk out of an inefficient cooling protogalaxy with high angular
momentum and/or lower mass \citep{sandage70,dalcanton97}.

Cosmological, numerical simulations of galaxy formation are challenged in forming bulgeless galaxies, known as the angular momentum problem (or angular
momentum catastrophe) \citep{navarro91}.~The simulated galaxies are too dense, too small, too centrally concentrated, and have lower angular momentum than observed 
because subhalos in a DM halo cool too fast, which causes angular-momentum loss by dynamical friction and 
merging of these clumps \citep{donghia04,piontek09a}. Feedback 
processes can suppress dramatic cooling and loss of angular momentum \citep{sommer03,okamoto05,robertson06,mayer08,scannapieco08}. 
Modern cosmological simulations show that it is possible to form exponential disk galaxies that are comparable to observations by 
using realistic models of feedback \citep{mayer08,piontek09b,governato10}. However, other studies claim that neither different 
kinds of feedback \citep{donghia04,donghia06} nor increased numerical 
resolution \citep{koeckert07,piontek09a} can resolve the angular momentum problem completely. Therefore, the formation 
of simple disk galaxies in a cosmological
framework is not yet well understood \citep{burkert08,mayer08}, and a detailed understanding of this topic is just at the beginning.

\section{Evolution}\label{evolution}

In the current $\Lambda$ cold DM ($\Lambda$CDM) framework of structure formation and evolution, 
galaxies in DM halos grow hierarchically by the absorption of
smaller substructures in sub halos \citep{searle78,white78,blumenthal84}. This means that disk galaxies have always been subject to merging and interaction.~Almost all
galaxies with present halo mass comparable to the Milky Way ($M_{DM}\sim10^{13}M_{\sun}$, $M_{stars}\sim10^{11}M_{\sun}$, \citet{dutton09}) are believed to have experienced a major merger 
(i.e., a merger with a
similar mass partner) \citep{stewart08,wang08,stewart09}.\footnote{Strictly speaking, the model of hierarchical clustering (HCM) implies that every structure has experienced a  
major merger when its history is followed far enough in the past \citep{khochfar01}.} Major mergers cause dramatic morphological transformations of disk galaxies. 
At the upper limit, a merger may cause the total destruction of the disk and the formation of a spheroidal, elliptical galaxy \citep[e.g.,][]{toomre77,barnes92,gardner01,cox08a}. 
Massive disks can then be rebuilt from gas deposited in a gas-rich (major) merger \citep{hammer09,robertson09,yang09} supported by the additional 
accretion of cold gas \citep{dekel06}. However, these so-called rebuilt scenarios assume that disks will be reformed around preexisting spheroidal bulges \citep{steinmetz03,springel05}. 
In less violent cases of major mergers, spheroidal bulges can formed by dynamically heated disk stars and accreted material \citep{aceves06,bournaud07,khochfar09}. 
In addition, new bulge stars can be formed from disk gas that lost its angular momentum by non-axisymmetric distortions due to galaxy-galaxy interactions \citep{noguchi01,benson04,hopkins09a,koda09}. 

Simple disk galaxies are low-mass systems (comparable to \object{M33} with $M_{DM}\sim10^{11.5}M_{\sun}$, $M_{stars}\sim10^{10}M_{\sun}$, \citet{dutton09}) 
that are not subject to frequent major merging events \citep{stewart08,wang08}. However, multiple minor mergers (with partners of mass ratios $\leq \frac{1}{3}$) 
are common for low massive galaxies at low redshifts \citep{bournaud07,bullock08,jogee09,stewart09}. Minor mergers heat the thin 
disks \citep[e.g.,][]{bullock08,kazantzidis08,kazantzidis09,purcell09}, let bulges grow \citep[e.g.,][]{naab03,donghia06,bournaud07,khochfar09} and could also form an 
elliptical galaxy \citep{bournaud07,combes09}. According to these model predictions, not many simple disks should have survived the cosmological evolution. 
Several recent studies \citep{robertson06,mayer08,hopkins09a,hopkins09b,koda09,weinzirl09} targeted 
this challenge and found that low-mass and gas-rich disk galaxies---such as simple disks---in combination with feedback processes 
can prevent substantial damage during mergers. In these models, the large amount of collisional gas suppresses violent relaxation of 
the angular momentum in the merger and subsequently conserves the disk structure of these galaxies. 

However, several observations of simple disk galaxies show signatures of galaxy-galaxy interactions. For example, many simple 
disk galaxies show warps, a possible indicator of ongoing morphological transformations \citep{reshetnikov95,uson03,matthews04}. 
It is also observed that some simple disks host a faint and diffuse thick stellar disk component \citep[][see also Section~\ref{structures}]{yoachim06}. 
These thick disks can be formed during merging events and contain large fractions of the stellar mass in such galaxies \citep{yoachim08}.

Another explanation of the observed frequency of simple disks is that they are exceptionally stable. Massive spheroidal components like bulges and DM halos can stabilize disks 
against external influence \citep{samland03,sotnikova06,kazantzidis09}. Simple disks have dominant, non-baryonic 
DM halos, see Section~\ref{properties}.\footnote{The structure of simple disks is determined by a rotation-supported, cold extended stellar disk; a dominant, 
spheroidal, non-baryonic DM halo (see Section~\ref{properties}); and no bulge. On the opposite end of
the morphological spectrum are the ellipticals with a dominant, hot stellar spheroid but almost without a cold disk and DM halo \citep{napolitano09}.}
Moreover, the disk thickness \citep{karachentsev97} of flat galaxies also is related to the dark halo. \citet{zasov02}, \citet{kregel05}, and \citet{mosenkov09} used samples of edge-on
disk galaxies including simple disks and found a correlation of the relative thickness of a stellar disk and the relative mass of the spheroidal component
including the DM halo. 
Nevertheless, disk galaxy evolution remains hotly debated and future papers will contain exciting insights in this field.

Disk galaxies can be also transformed via internal, secular evolution. We are currently in a cosmological transition era where secular evolution is becoming an
important process \citep{kormendy05}. Non-axisymmetric structures like bars and  
oval disks support internal disk instabilities and transport gaseous material to the disk center (\citet{kormendy83,kormendy04}, see also \citet{combes90}). 
Subsequent central star formation forms a pseudobulge 
with disk-like properties such as disky isophotes when seen edge-on, exponential surface brightness profiles, and low velocity dispersion \citep{kormendy05,fisher08}. Bars are frequently detected 
in bulgeless galaxies \citep{matthews97,barazza08}, making simple disks potential candidates for secular evolution. 
Flat and superthin galaxies are ideal for studying the predictions of secular evolution and the growth 
of pseudobulges because we do not know how many low-mass disks are affected by this internal evolution and if it is a common phenomenon in these objects. 

\section{Fractions of simple disks}

Bulgeless simple disk galaxies are common in the local Universe \citep{matthews97,boeker02,goto03,barazza08,cameron09}.~The first comprehensive search for disk-dominated and bulgeless galaxies was initiated 
by \citet{karachentsev89} in order to map cosmic flows. Karachentsev used a simple but effective method to classify these galaxies by selecting only edge-on disks where 
bulges can be easily detected and the vertical structure can be studied. This work resulted 
in the ``Flat Galaxy Catalog'' \citep[FGC,][]{karachentsev93} and the ``Revised Flat Galaxy Catalog'' 
\citep[RFGC,][]{karachentsev99a}. These optical all-sky surveys are supplemented by the near-infrared ``The 2MASS-selected Flat Galaxy Catalog'' \citep{mitronova04}. 
Follow-up optical and HI radio observations for FGC and RFGC galaxies are collected in \citet{giovanelli97}, \citet{dalcanton00}, \citet{matthews00}, \citet{makarov01}, 
\citet{mitronova05}, and \citet{huchtmeier05}.

\citet{kautsch06a} used the first data release of the Sloan Digital Sky Survey \citep[SDSS DR1,][]{abazajian03} in order to collect a uniform and homogeneous
catalog of edge-on disk galaxies. Similar to the FGC and RFGC, \citet{kautsch06a,kautsch06b} selected the objects based on axial ratio ($\frac{a}{b} > 3$), angular
diameter ($a > 30 \arcsec$), and apparent magnitude ($m < 20$ mag in the SDSS $g$ band) within a certain color range. The galaxies were then separated into a
morphological sequence ranging from objects with bulges to bulgeless simple disks and 
irregulars: Sa(f), Sb(f), Sc(f), Scd(f), Sd(f) and Irr(f); (f) indicates that the galaxies contain flat disks seen edge-on. This
automated classification is based on bulge size and disk flatness. The bulge size is represented by the light concentration index in the SDSS $r$ band. 
This concentration index is the ratio of the
Petrosian radii given in the SDSS for each object that contains 90\% and 50\% of the Petrosian flux in the same band, 
respectively \citep[see][for the definition of the Petrosian parameters in the SDSS]{stoughton02}. The disk flatness parameter, $e$, 
is the luminosity weighted
mean ellipticity of the elliptical isophotes in the SDSS $r$ band fitted to each 
catalog galaxy \citep{kautsch06a}.\footnote{\citet{kautsch09a} found that $e$ can also be directly derived from the image moments available in the SDSS archive.} 

The fraction of the simple disk class Sd(f) is 16\% among the disk galaxies in the \citet{kautsch06a} catalog. This fraction increases to 32\% if the 
seemingly bulgeless (but less strictly defined) Scd(f) types are included. 
\citet{kautsch09a,kautsch09b} confirmed these fractions by using the SDSS DR6 \citep{adelman08}. \citet{kautsch09a} 
also compared the fraction of simple disks in the local Universe with other recent studies \citep{karachentsev99a,karachentsev04,allen06,kautsch06a,barazza08,koda09} and found 
a simple disk fraction of 16$\pm$3\% on average among disk galaxies. This frequency shows that bulgeless galaxies comprise a non-negligible fraction of spiral galaxy systems. 
It is possible that small and compact bulges are obscured due to dust extinction \citep{tuffs04,driver08}. However, this is unlikely in a majority of simple disks because they are
observed to be transparent and these bulges would have different properties to those of classical bulges predicted by theoretical models \citep{cameron09}. 
Field studies show that the number density of large bulgeless galaxies is constant (maybe slightly increasing) at redshifts $0\leq z \leq 1$ whereas the number of galaxies with bulges 
decreases at larger distances \citep{sargent07,dominguez09}. 

Bulgeless galaxies are located in all environments, ranging from low to high density \citep{kautsch05,kautsch09c}. The majority of these galaxies are weakly associated with galaxy 
clusters and can be found in more isolated environments comparable to galaxy groups and the field \citep{kudrya97,karachentsev99b,kautsch09c}. Because of the 
low relative velocities of group galaxies, merging and morphological processes that transform late-type galaxies into bulge-dominated and spheroidal galaxies 
are common in the group environment \citep[e.g.,][]{barnes85,kautsch08,tran08}. 
This implies that simple disks either have to be stable against morphological preprocessing or are located in this environment due to recent infall. 

\section{Global Properties of Simple Disks}\label{properties}

Simple disks are not a separate morphological class, but rather at the end of a smooth continuum without a well-defined 
boundary \citep{matthews97,kautsch06a}. The continuum ranges from massive, stellar disk galaxies with substantial bulges and with 
high surface brightnesses to the bulgeless galaxies with lower masses and surface brightnesses 
\citep[e.g.,][]{schombert92,karachentsev93,matthews99,dutton09,ganda09}. 

Figures 2$-$6 illustrate these properties for different edge-on galaxies. 
Two prototypical superthin galaxies, \object{UGC 07321} and \object{IC 2233},\footnote{\object{IC 2233} = \object{UGC 04278}} are highlighted with large cross symbols. 
The objects in all figures represent a randomly selected subsample (in order to avoid making the plots too crowded) from the SDSS DR6 edge-on disk galaxy collection by \citet{kautsch09a}. 
These galaxies are matched with the \citet{giovanelli97} and \citet{huchtmeier05} catalogs in order to obtain their rotational velocities, except for UGC 07321 and IC 2233 for which the 
kinematic information was
collected from \citet{matthews99} and \citet{matthews08a}, respectively. The numbers of the objects slightly vary between the diagrams because some SDSS parameters or kinematics 
are not provided for every
individual galaxy.  

\begin{figure}
\epsscale{1}
\plotone{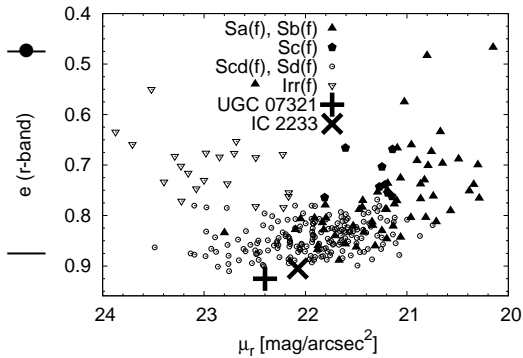}
\caption{This figure shows the total surface brightness, $\mu$, versus the disk flatness, $e$, for a random sample of the \citet{kautsch09a} edge-on galaxies. 
See text for a description of the parameters. The symbols above and below the y-axis label illustrate the increasing bulge size of the galaxies along this axis. 
The bulgeless galaxies, Scd(f), Sd(f), and Irr(f), exhibit lower total 
surface brightnesses compared to the disk galaxies with bulges.}\label{fig2}
\end{figure}

\begin{figure}
\epsscale{1}
\plotone{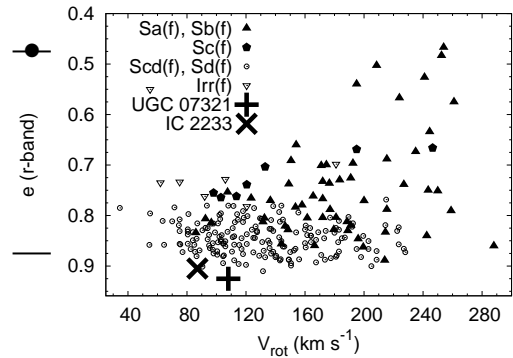}
\caption{This figure shows the rotational velocity, $v_{rot}$, versus the disk flatness, $e$, for the same random sample of Fig.~\ref{fig2}. 
Galaxies with bulges rotate faster on average compared to disk-dominated and bulgeless systems.}\label{fig3}
\end{figure}

\begin{figure}
\epsscale{1}
\plotone{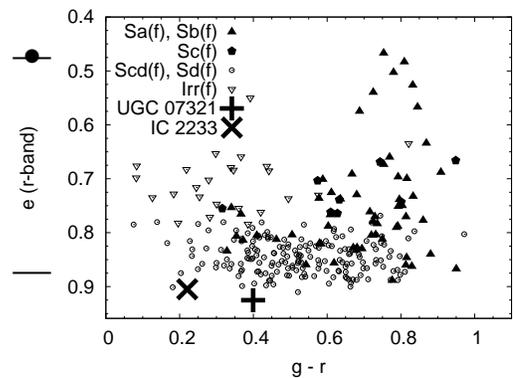}
\caption{This figure shows the $g-r$ color versus the disk flatness, $e$, for the same random sample of Fig.~\ref{fig2}. Galaxies 
with bulges are redder on average compared to disk-dominated and bulgeless systems.}\label{fig4}
\end{figure}

Figure~\ref{fig2} shows the lower total surface brightnesses of simple disks compared to disks with bulges. 
However, this does not mean that every simple disk is an LSB galaxy \citep{kautsch06a} nor that every LSB galaxy is bulgeless \citep{bizyaev04}. 
The previously discussed flatness parameter $e$ in the Figures \ref{fig2}$-$\ref{fig4} is derived from the SDSS image moments as shown in \citet{kautsch09a}. 
The SDSS does not contain image moments for IC 2233 because of a nearby, saturated projected star, therefore I use its isophotal ellipticity as given in the archive as a proxy for $e$. 
The total surface brightness of each galaxy in Figure~\ref{fig2} is derived by using the 
parameters $\mu$ = \texttt{petroMag}+\texttt{rho}, which are given in the SDSS archive \citep{stoughton02}. 
\texttt{rho} is five times the logarithm of the Petrosian radius. No correction for inclination and extinction in the individual objects 
is applied.

On average, bulgeless disks rotate slower than galaxies with bulges as shown in Figure~\ref{fig3}. 
The HI line width at 50\% peak flux ($W_{50,c}$) from \citet{giovanelli97} and \citet{huchtmeier05} is used to derive the 
rotational velocities for the sample galaxies \citep[$v_{rot}$ = $W_{50,c}/2$,][]{dalcanton02}. Considering the rotational velocity as a proxy for the total mass of the objects, the
figure implies that flat galaxies are low-mass systems. 

Only a few edge-on simple disks have been studied in detail so far. Therefore I will focus on studies of LSB superthin simple disks 
such as \object{UGC 07321} and \object{IC 2233} in 
this and the next section. 
The results from these prototypes are presumably valid for most of the simple disks.
Generally, simple disks have low metallicities and blue global colors \citep{matthews97,matthews08a,cameron09} which 
places them in the blue cloud of galaxies \citep{strateva01,baldry04}. Figure~\ref{fig4} shows the apparent colors for different edge-on disk galaxy types. The 
colors are derived from the Galactic extinction corrected Petrosian $g$ and $r$ magnitudes from the SDSS archive \citep{stoughton02}. Although no inclination
correction is applied, edge-on galaxies with bulges appear to be redder compared to the average color of simple disks. The color range of bulgeless disks can represent
variations of the recent star-formation rates \citep{lee09,west09}. Variations in metallicities and reddening due to different dust content also can cause differences in 
colors, but these effects are considered to be small because simple disks host only small amounts of interstellar dust, discussed later in this section.  

Simple disks are not necessarily young. Many contain old stellar populations \citep[$\leq$ 10 Gyrs,][]{bergvall95,deblok95}. \citet{matthews99} and 
\citet{matthews08a} also find radial color gradients in edge-on superthins with a central mix of stellar populations of 
different ages and a very young population in the outer disk. Additionally, they find a population of redder, older stars at higher scaleheights. This suggests that the objects formed slowly in time from the inside out and experienced vertical dynamical heating. 

These studies also show signatures of ongoing, localized star formation such as HII regions, OB associations, and 
candidate supergiant populations \citep{bergvall95,matthews99,matthews08a}. The global star formation rates of the prototypical superthins are: UGC 07321,
SFR$_{IRAS} \sim 0.006$ $M_{\sun} yr^{-1}$ \citep{matthews03,uson03}; and IC 2233, SFR$_{IRAS} \sim 0.02$ $M_{\sun} yr^{-1}$ \citep{matthews08a}. These estimates are
at the low end of observed star formation rates for Sd spirals \citep{kewley02}. Therefore, Matthews and coworkers conclude that the superthin galaxies are
underevolved systems in the sense of star formation \citep[e.g.,][]{matthews99,matthews08a}. The low global star formation rates
can be explained by the HI surface density being too low to efficiently form stars \citep[e.g.,][]{vdhulst93,schombert01} and a high velocity dispersion of the gas 
that makes the disks stable against star formation \citep{banerjee10}. Interesting future work could be done concerning the star 
formation rate per area, or star-formation rates, normalized to the physical sizes of the galaxies \citep[cf.,][]{hunter04}. The specific star formation rates of the superthins might be
also higher than the global star formation when considering the low stellar masses \citep[e.g.,][]{dutton09} of these bulgeless systems. 

Bulgeless galaxies contain large amounts of atomic, neutral HI gas \citep{karachentsev99c,matthews00,makarov01,matthews08a}. 
The gas is extended throughout the stellar disk \citep{matthews99,matthews08a}.~Also hot, ionized HII gas---distributed in clumps---exists in 
simple disks \citep{matthews99}. Molecular H$_{2}$ gas 
as traced by carbon monoxide, CO, is weakly detected in late-type spirals and edge-on simple disks \citep{young89,matthews01c,boeker03b,matthews05}.

\begin{figure}
\epsscale{1}
\plotone{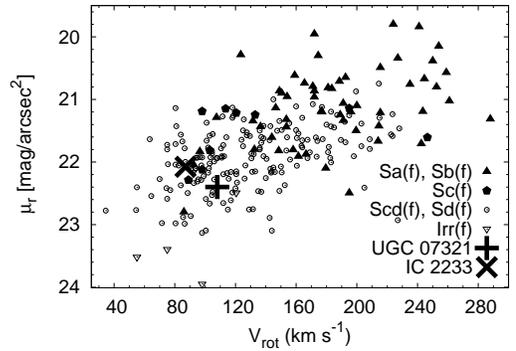}
\caption{This figure shows the rotational velocity, $v_{rot}$, versus total surface brightness, $\mu$, for the same random sample of Fig.~\ref{fig2}. 
Bulgeless and disk-dominated systems rotate slower and have lower surface brightnesses compared to galaxies with bulges.}\label{fig5}
\end{figure}

\begin{figure}
\epsscale{1}
\plotone{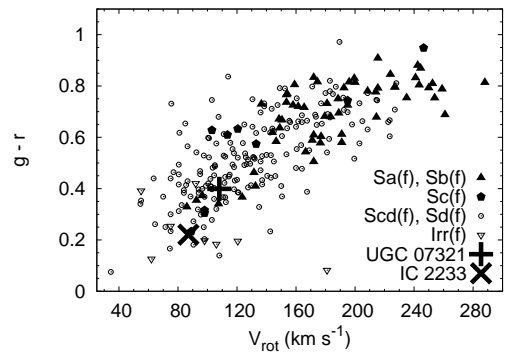}
\caption{This figure shows the rotational velocity, $v_{rot}$, versus the $g-r$ color for the same random sample of Fig.~\ref{fig2}. 
Bulgeless and disk-dominated systems have bluer colors and are slow rotators compared to galaxies with bulges.}\label{fig6}
\end{figure}

The amount of dust is generally low \citep{matthews01w,stevens05} as implied by the transparency of the edge-on 
simple disks \citep{matthews99,matthews01w,karachentsev02,matthews08b}. 
In contrast to the organized dust lanes in edge-on spiral galaxies with bulges, simple disks
show a clumpy and diffuse distribution of dust \citep{matthews99,matthews00b,matthews01w}. \citet{dalcanton04} found that organized dust lanes 
appear only in edge-on galaxies with bulges and relative fast rotational velocities. These authors suggest that the galaxies with organized dust lanes are more 
gravitationally unstable, which leads to fragmentation and gravitational collapse
along spiral arms and subsequently smaller gas scaleheights, pronounced dust lanes, star formation, and high surface brightnesses. 
In contrast, the distribution of dust in edge-on simple disks is clumpy if their rotational velocity is below $v_{rot} = 120$ km s$^{-1}$. In this case the dust has not
settled into a thin lane and therefore appears patchy and diffuse because the simple disks are gravitationally stable and have low star-formation rates, 
which also implies lower metallicities and lower mass. 
This explains the lower total surface brightnesses and slower rotation of simple disk galaxies compared to 
galaxies with bulges as shown
in Figure~\ref{fig5}. The ideas from \citet{dalcanton04} are then also visible in Figure~\ref{fig6}: more massive galaxies as indicated by their larger rotational velocities form 
more stars at earlier times and have redder
present day colors. In contrast, slow rotators tend to have thicker gas disks and thus less efficient star formation and therefore spread their star formation out over a longer
time \citep[see also][]{banerjee10}. In
this way they can remain blue for longer times.

According to the ideas in \citet{dalcanton04}, Figure~\ref{fig6} suggests that the rotational velocity 
(determined by the baryonic mass and DM) regulates the average star formation
history \citep[cf.,][]{kennicutt98}. While bulges become more common at large $v_{rot}$, their presence is not necessarily related to a red galaxy color 
because bulgeless galaxies can also have red colors. 
In other words, if a galaxy has a high rotational velocity, it forms stars quickly \citep{dalcanton04} and so has red colors; and it is also 
more likely to produce a
bulge. Both characteristics are tied to the rotational velocity but it remains unclear whether a red color and the presence of a bulge are correlated independent of $v_{rot}$. 
The lower surface brightnesses of simple disks indicate that the probability for bulge
formation depends on the mass and $v_{rot}$ of the host galaxy. As suggested by \citet{kautsch06a}, this correlation can be linked to the models where internal, 
secular disk instabilities are responsible for forming bulges through the dependence 
of the Toomre Q-parameter on disk surface density \citep[e.g.,][]{immeli04}. 

The rotation curves of edge-on simple disk galaxies are generally flat and slowly rising throughout their stellar disk \citep[][]{matthews99,mendelowitz00,makarov01,vdkruit01,zackrisson06}, see also \citet{zasov03}.
These solid-body rotation curves are typical for late-type irregular galaxies, making simple disks the simplest dynamical type of disk galaxies.~The 
rotation curves and axial ratios indicate that these galaxies are completely DM dominated, even in their centers, and are
surrounded by a spherical dark halo \citep{karachentsev91,mendelowitz00,zasov02,uson03,zackrisson06,banerjee10}. 
These rotation curves are also useful to probe DM profiles in disk galaxies. 
Numerical N-body simulations of $\Lambda$CDM predict central dark halo mass densities significantly larger and cuspier 
than observed in LSB simple disks \citep[``core/cusp problem,'' see][and references therein]{navarro97}. In contrast to the models, the observations 
show nearly constant density cores \citep{zackrisson06,mcgaugh07,kuziodenaray09}. 

The Tully-Fisher relation of edge-on simple disks can be used for 
estimations of distances, luminosities, diameters, and other parameters
\citep[e.g.,][]{karachentsev91b,kudrya97,karachentsev99c,karachentsev02}. 
The dust corrected Tully-Fisher relation for faint and bulgeless LSB galaxies indicates that their absolute magnitudes appear to be fainter than for spirals 
with bulges for a fixed HI line width, i.e., faint LSB simple disks rotate faster for a predicted luminosity \citep{kudrya97,matthews01w}.

The far end of the continuous sequence of properties is occupied with bulgeless irregulars, see Figures~\ref{fig2}$-$\ref{fig6}. The main difference is their thicker appearance. 
The underlying
reason for this structural difference may be of kinematical origin, where turbulent motion can compete with ordered rotation 
because of low rotational velocities in irregulars \citep{seiden79,sung98}.  
This in turn leads to low surface brightnesses because more stellar
material is distributed over a wide range of disk scale heights, which produces low stellar surface densities \citep{schombert06}. 
This trend is visible in Figure~\ref{fig5}, although the sample of irregulars is small in the present study 
because of the lack of available kinematic information for the presented objects. \citet{schombert06} concludes that the random gas motion leads to stochastic and slow star formation
compared to coherent patterns of star formation in flatter disks \citep[see also][]{banerjee10}. However, the difference between flat disks and puffy irregulars is not understood in
detail, considering that Irr(f) objects and flat disks can have similar values of their rotational velocity (Fig.~\ref{fig3}). 

\section{Structures in Simple Disks}\label{structures}

The radial surface brightness profiles of edge-on simple disks are close to projected exponentials, as they are also for simple disks at other, less inclined viewing angles \citep{matthews99,boeker03,dutton09}. The profiles in some bulgeless LSB galaxies decrease from an exponential fit in the
central regions but it is unknown whether a strong DM dominance in the centers of the galaxies is responsible for the deficit of the stellar densities
\citep{zackrisson06}.

The vertical stellar structure of edge-on simple disks can be fit with a variety of profiles (isothermal sech$^2$, sech, or exponential profiles) which differ only at small 
heights \citep[][and references therein]{matthews00b}.~At large scaleheights, single profiles sometimes deviate from one-component fits which could be explained by a second, thick stellar 
disk component \citep{yoachim06}. These thick disks appear to be older than the thin disk and have distinct, slower kinematics, even counterrotation 
\citep{matthews00b,mould05,yoachim08}.~Whereas internal or external 
heating via dynamical friction can be responsible for the thick disk, current studies favor direct accretion of the 
thick-disk material during minor mergers \citep{yoachim08}. The sample studied so far is very small. Using larger samples, edge-on simple disks may be an excellent tool to test the
different thick-disk formation theories. 

\citet{ferrarese00} and \citet{gebhardt00} (see also \citet{ganda09}) found a tight relation between the mass of supermassive black holes and the velocity dispersion of bulges in disk 
galaxies.~This
relation can be explained in a hierarchical universe where bulge and black hole growth is a consequence of merging galaxies \citep{peng07,wang08}. 
In this respect, simple disk galaxies are expected to be
black hole free. However, there is growing evidence that this is not always true; for example: \object{NGC 1042}, \citet{shields08}; \object{NGC 3621}, \citet{satyapal07}, \citep{gliozzi09}; or 
\object{NGC 4395}, \citet{filippenko03}, among other simple disks. 
Nuclear star clusters are also often found in bulgeless galaxies \citep{boeker02,walcher05,walcher06,rossa06}.

\section{Outlook}
We need studies focused on edge-on bulgeless galaxies independent of their surface brightnesses and other
selection criteria to investigate the properties and the formation and evolution of simple disks. Large surveys already contain much of the needed material for such future investigations. 
Interesting work could be performed by
tracing the frequency and properties of simple disks in different environments and redshifts in order to paint a picture of their evolution history and progenitor systems
\citep[cf.,][]{elmegreen04}. A census about the total halo, stellar, and gas masses would shed light on the stability of the disks and eventually on the mystery of the dark matter. 

Simple disk galaxies are ideal objects to test current cosmological theories of the formation, evolution, and morphological 
transformations of galaxies; to explore unknown properties of these objects; and to fascinate people in the International Year of Astronomy and beyond.


\acknowledgments

The author thanks the anonymous referee for very helpful comments and suggestions, and Eva Grebel, John Gallagher, Anthony Gonzalez, 
Fabio Barazza, Kent Cueman, Ata Sarajedini, and Jakob Walcher for critical reading of the manuscript and enlightening discussions about this topic. This research has 
made use of the VizieR catalogue access tool, CDS, Strasbourg, France, and the NASA/IPAC Extragalactic Database (NED) which is operated by the Jet Propulsion Laboratory, 
California Institute of 
Technology, under contract with the National Aeronautics and Space Administration. Funding for the SDSS and SDSS-II has been provided by the Alfred P. Sloan Foundation, the Participating Institutions, the National Science Foundation, the U.S. Department of Energy, the National Aeronautics and Space Administration, the Japanese Monbukagakusho, the Max Planck Society, and the Higher Education Funding Council for England. The SDSS Web Site is http://www.sdss.org/.

The SDSS is managed by the Astrophysical Research Consortium for the Participating Institutions. The Participating Institutions are the American Museum of Natural History, Astrophysical Institute Potsdam, University of Basel, University of Cambridge, Case Western Reserve University, University of Chicago, Drexel University, Fermilab, the Institute for Advanced Study, the Japan Participation Group, Johns Hopkins University, the Joint Institute for Nuclear Astrophysics, the Kavli Institute for Particle Astrophysics and Cosmology, the Korean Scientist Group, the Chinese Academy of Sciences (LAMOST), Los Alamos National Laboratory, the Max-Planck-Institute for Astronomy (MPIA), the Max-Planck-Institute for Astrophysics (MPA), New Mexico State University, Ohio State University, University of Pittsburgh, University of Portsmouth, Princeton University, the United States Naval Observatory, and the University of Washington.


\bibliography{refs}

\end{document}